\documentclass[prb,showpacs,amsmath,amssymb,superscriptaddress,twocolumn,letterpaper]{revtex4}

\usepackage{amssymb}
\usepackage{graphicx}
\usepackage{color}
\usepackage{dcolumn}
\usepackage{bm}
\usepackage{multirow}

\begin{document}
\newcommand{\rte}{$R$Te$_3$}
\newcommand{\ce}{CeTe$_3$}
\newcommand{\gd}{GdTe$_3$}
\newcommand{\tb}{TbTe$_3$}
\newcommand{\dy}{DyTe$_3$}
\newcommand{\tupper}{$T_{\mathrm{CDW,1}}$}
\newcommand{\tlower}{$T_{\mathrm{CDW,2}}$}
\newcommand{\tc}{$T_c$}
\newcommand{\pc}{$P_c$}
\newcommand{\tn}{$T_{\mathrm{N}}$}
\newcommand{\hc}{$H_{c2}(T)$}

\abovedisplayskip=5pt
%\abovedisplayshortskip=5pt
\belowdisplayskip=5pt
%\belowdisplayshortskip=5pt

\title{Pressure dependence of the charge-density-wave and superconducting states in \gd, \tb\ and \dy}

\author{D.\ A.\ Zocco}
%\altaffiliation[Present address: ]{Institute of Solid State Physics (IFP), Karlsruhe Institute of Technology, D-76021 Karlsruhe, Germany.}
\email{diego.zocco@kit.edu}
\affiliation{Department of Physics, University of California, San Diego, La Jolla, CA 92093, USA}
\affiliation{Institute of Solid State Physics (IFP), Karlsruhe Institute of Technology, D-76021 Karlsruhe, Germany}
\author{J. J. Hamlin}
%\affiliation{Department of Physics, University of California, San Diego, La Jolla, CA 92093, USA}
\altaffiliation[Present address: ]{Department of Physics, University of Florida, Gainesville, FL 32611, USA}
\affiliation{Department of Physics, University of California, San Diego, La Jolla, CA 92093, USA}
\author{K.\ Grube}
\affiliation{Institute of Solid State Physics (IFP), Karlsruhe Institute of Technology, D-76021 Karlsruhe, Germany}
\author{J.-H. Chu}
\author{H.-H. Kuo}
\author{I. R. Fisher}
\affiliation{Department of Applied Physics, Geballe Laboratory for Advanced Materials, Stanford University, CA 94305, USA}
\author{M. B. Maple}
\email{mbmaple@ucsd.edu}
\affiliation{Department of Physics, University of California, San Diego, La Jolla, CA 92093, USA}

\begin{abstract}
We present electrical resistivity and ac-susceptibility measurements of \gd, \tb\ and \dy\ performed under pressure. An upper charge-density-wave (CDW) is suppressed at a rate of $\mathrm{d}T_{\mathrm{CDW,1}}/\mathrm{d}P$\,$\sim$\,$-$85\,K/GPa. For \tb\ and \dy, a second CDW below \tlower\ increases with pressure until it reaches the \tupper($P$) line. For \gd, the lower CDW emerges as pressure is increased above $\sim$\,1\,GPa. As these two CDW states are suppressed with pressure, superconductivity (SC) appears in the three compounds at lower temperatures. Ac-susceptibility experiments performed on \tb\ provide compelling evidence for bulk SC in the low-pressure region of the phase diagram. We provide measurements of superconducting critical fields and discuss the origin of a high-pressure superconducting phase occurring above 5\,GPa.
\end{abstract}


\maketitle

\section{Introduction}\label{section1}

Superconductivity (SC) and charge-density-waves (CDW) are collective electronic phenomena that originate from electron-electron and electron-phonon interactions, and the concept of Fermi surface competition between these collective states is one of the most fundamental problems of condensed matter physics.\cite{wilson75,bmcm76,balseiro79} Materials that present the interplay between SC and CDW are characterized by reduced dimensionality of their electronic and structural properties, and include, for example, the single element actinide $\alpha$-uranium,\cite{lander94} one-dimensional organic chains,\cite{kondo10} layered transition-metal chalcogenides,\cite{berthier76,briggs80,sipos08,kusmartseva09} and the copper-oxide high-temperature superconductors.\cite{kivelson03} Recently, the family of rare-earth tritellurides (\rte) compounds has been added to this list, when SC was found in \tb\ upon the suppression of a CDW state with the application of pressure.\cite{hamlin09a} 

Rare-earth tritellurides are quasi-two-dimensional layered materials consisting of double planar layers of Te-Te atoms, separated by double corrugated slabs of rare-earth ($R$\,=\,La-Nd, Sm, and Gd-Tm) and tellurium atoms.\cite{dimasi95} The \rte\ compounds form in a weakly orthorhombic crystal structure (space group \textit{Cmcm}), with the long $b$ axis perpendicular to the Te planes.\cite{norling66} The elongated crystal structure is responsible for the weak hybridization between the Te and the $R$Te layers, reflected in large conductivity anisotropies of the order of 10$^3$ (Ref.\,\onlinecite{dimasi94}). A CDW forms in the Te-Te planes below \tupper\ with an incommensurate wavevector $\boldsymbol{q_1}$\,=\,(0, 0, $\sim$2/7$c^{*}$). For the heavier \rte, a second CDW forms below \tlower\,$<$\,\tupper, with $\boldsymbol{q_2}$\,=\,($\sim$1/3$a^{*}$, 0, 0).\cite{ru08a} The transition temperatures \tupper\ and \tlower\ can be tuned by adjusting the in-plane lattice parameters via chemical pressure\cite{dimasi95} and with the application of external pressure.\cite{sacchetti07a, sacchetti09a, zocco09a} Recently, a third type of CDW order has been suggested to appear in the \rte\ compounds, possibly arising from a lifting of the degeneracy of conduction bands of double Te sheets.\cite{hu14} Angle-resolved photoemission spectroscopy (ARPES) experiments confirm the quasi-2D nature of the electronic properties, showing partial gapping of the Fermi surface (FS) below the CDW ordering temperatures.\cite{dimasi95, gweon98, komoda04, laverock05, brouet08} Fermi surface nesting has been revealed as a strong candidate for driving the formation of CDWs in these compounds. On the other hand, recent inelastic x-ray scattering and Raman spectroscopy experiments showed compelling evidence that electron-phonon coupling is playing an important role in the charge ordering and the lattice distortion observed in these materials.\cite{johannes08, lazarevic11, eiter13, maschek14} 

So far, most of the work has focused on understanding the complex characteristics of the multiple CDW states. Studying the interplay of these CDWs with the magnetism arising from the $R$-Te sublattice and the recently discovered pressure-induced SC in \tb\ requires the performance of experiments at high pressures and below 10\,K. In this paper, we present high-pressure electrical resistivity and ac-magnetic susceptibility measurements on \gd, \tb\ and \dy. Similar to \tb, we found pressure-induced SC in \gd\ and \dy. We further studied the superconducting state of \tb\ using ac-susceptibility and resistivity, providing compelling evidence of bulk SC in the low-pressure region of the phase diagram of these materials.

\section{Experimental details}\label{section2}

Single crystals of \gd, \tb\ and \dy\ were grown in an excess of tellurium via a self flux technique.\cite{ru06a} Electrical resistivity measurements under hydrostatic pressures were performed employing a Cu-Be piston-cylinder cell using a 1:1 mixture of n-pentane:isoamyl alcohol as the pressure transmitting medium which remains hydrostatic below 3\,GPa. For pressures up to 16\,GPa, we utilized a Cu-Be Bridgman-anvil cell using solid steatite as the quasi-hydrostatic pressure medium. Pressure was changed at room temperature and determined with a lead or tin manometer.\cite{bireckoven88,smith69} Pressure gradients were inferred from the width of the manometer superconducting transition, being as large as 2\% and 10\% of the total pressure for the hydrostatic and the Bridgman-anvil cell experiments, respectively. In all cases, the electrical resistance in the $ac$-plane was measured using a 4-lead technique and a Linear Research Inc.\,LR-700 AC resistance bridge operating at 16\,Hz. Contacts of $\sim$\,2\,$\Omega$ were obtained after evaporating gold pads onto the surface of the samples, which also served as protection against the effect of air. Platinum or gold leads (50\,$\mu$m) were attached to the gold pads with five minute silver epoxy. Temperatures as low as 1.2\,K were attained in a conventional $^4$He Dewar connected to a pumping system and for temperatures ranging from 0.1\,K\,$\leq\,$T$\,\leq$\,2\,K, an Oxford Kelvinox MX100 $^{3}$He-$^{4}$He dilution refrigerator was utilized.

Ac-susceptibility measurements were performed in the hydrostatic cell. A set of secondary coils was designed to fit inside the 1/4''-diameter Teflon capsules, illustrated in Fig.\,\ref{fig04}(b). Each coil of the secondary contained $\sim$\,500 turns (19 layers) of AWG50-gauge Cu wire (bare nominal diameter of 25\,$\mu$m), for a total electrical resistance of 210\,$\Omega$. The coils built in the main body of the clamp were used as primary coils. For an excitation field of 5\,Oe and 1023\,Hz, an induced voltage of 100\,$\mu$V was estimated to arise from the shielding of a 100\,$\mu$m-thick type-I superconducting sample that would fit in the 1\,mm-diameter space of the spools. The induced signal was detected using a Linear Research Inc.\,LR-700 AC resistance bridge or an Stanford Research SR-830 lock-in amplifier.

\section{Results}\label{section3}

\subsection{Electrical resistivity ($H$\,=\,0)}\label{section3.A}
\begin{figure}[t]
\begin{center}
\includegraphics[width=3.0in]{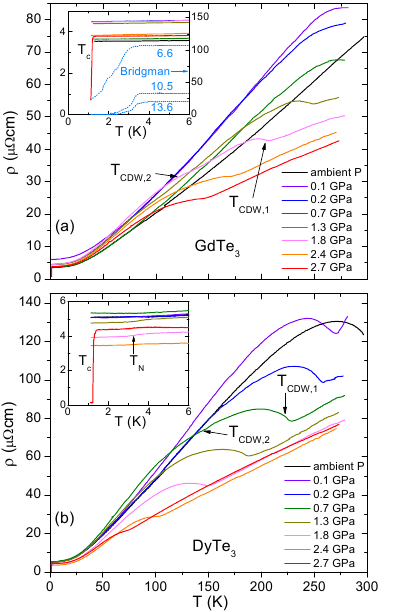}
\caption{(Color online) In-plane electrical resistivity $\rho$ versus temperature $T$ of (a) \gd\ and (b) \dy\ at different pressures (hydrostatic cell). CDW ordering temperatures \tupper\ and \tlower\ were obtained from the derivatives of these curves (not shown). Insets display the superconducting and magnetic ordering features occurring at low temperatures. Superconducting transitions in \gd\ obtained from the Bridgman-anvil cell are shown in the inset of (a).}
\label{fig01}
\end{center}
\end{figure}
Figures\,\ref{fig01}(a) and (b) show the electrical resistivity $\rho(T)$ of \gd\ and \dy, respectively, measured in the hydrostatic cell up to 2.7\,GPa and down to 1.2\,K. For \gd, the curves present a shoulder above 250\,K for pressures of 0.1, 0.2 and 0.7\,GPa, which progresses as the clear onset to the upper CDW at \tupper\ at higher pressures. Pressure shifts \tupper\ towards lower temperatures, reaching a value of about 150\,K at 2.7\,GPa. At 1.3, 1.8 and 2.4\,GPa, weaker features near 100, 130 and 140\,K, respectively, indicate the onset to the lower CDW at \tlower. For \dy, \tupper\ appears at 0.1\,GPa, and it is also suppressed with pressure, reaching 77\,K at 2.7\,GPa. For this sample, the lower CDW can be determined only for the 0.2 and 0.7\,GPa measurements, with \tlower\,$\sim$\,95 and 165\,K, respectively. The combination of the unidirectional character of both CDWs and an anisotropic conductivity in the $a$-$c$ plane could be responsible for the relatively weak signatures observed at $T_{\mathrm{CDW,1}}$ and $T_{\mathrm{CDW,2}}$.\cite{sinchenko14} These data are summarized in Fig.\,\ref{fig05} and discussed in Section\,\ref{section4}.
\begin{figure*}[t]
\begin{center}
\includegraphics[width=1.0\textwidth]{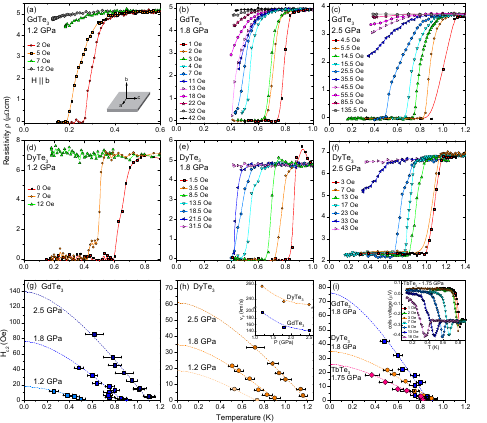}\\
\caption{(Color online) Determination of critical fields of \gd, \tb\ and \dy\ at different pressures obtained in the hydrostatic cell experiments. Magnetic fields were applied perpendicular to the planes; \textit{i.\,e.}, $H$\,$\parallel$\,$b$-axis. (a-c) Electrical resistivity of \gd. (d-f) Electrical resistivity of \dy. (g) \hc\ of \gd\ at 1.2, 1.8 and 2.5\,GPa. (h) \hc\ of \dy\ at 1.2, 1.8 and 2.5\,GPa. The inset shows the evolution of the Fermi velocities $v_F$ obtained from the slopes near \tc. (i) \hc\ of \gd, \tb\ and \dy\ at similar pressures. For \tb, data were obtained from the ac-susceptibility measurements displayed in the inset. The dashed lines plotted in (g-i) correspond to calculations of \hc\ with the WHH model.}
\label{fig02}
\end{center}
\end{figure*}

The localized magnetic moments of the rare-earth atoms order antiferromagnetically at ambient pressure, with \tn\,=\,9.7\,K and 11.3\,K for \gd\ and \tn\,=3.44\,K and 3.6\,K for \dy.\cite{ru08b} For \gd, it is not possible to observe the AFM transitions for any of the applied pressures. In the inset of Fig.\,\ref{fig01}(b), the onset to the AFM order of \dy\ at \tn\ appears as a low step below 4\,K, similar to \tb.\cite{hamlin09a}

At 2.7\,GPa, both \gd\ and \dy\ display a sharp feature near 1.3\,K and 1.45\,K, respectively [insets of Fig.\,\ref{fig01}(a) and (b)], indicating that pressure induces SC in these compounds, similar to \tb.\cite{hamlin09a} For \dy, a 100\,mK--wide transition to zero resistance is observed. The inset of Fig.\,\ref{fig01}(a) also displays data at 6.6, 10.5 and 13.6\,GPa obtained in a Bridgman-anvil cell for \gd\ (light-blue dotted lines, right axis). Full superconducting transitions to zero resistance are observed above 3\,K for the highest pressure,  although in this case large pressure gradients result in wider transitions.

\subsection{Critical fields}\label{section3.B}
Figures \ref{fig02}(a\,--\,f) present the electrical resistivity of \gd\ and \dy\ obtained in a $^3$He--$^4$He dilution refrigerator for three different pressures (2.5, 1.8 and 1.2\,GPa) accessed by decreasing the applied load to the clamp at room temperature. Magnetic fields smaller or of the order of 100\,Oe applied parallel to the out-of-plane $b$-axis of the samples were sufficient to suppress the superconducting state.\cite{note1} Small values of critical fields of the order of 200\,Oe were also observed in the layered pressure-induced superconductor 1$T$-TiSe$_2$.\cite{kusmartseva09} Values of \tc\ were determined at the onset of the transitions. Figures \ref{fig02}(g) and 2(h) display the critical fields for \gd\ and \dy\, respectively. Our data can be described well with the solutions to the linearized-Gor'kov equations developed by Werthamer, Helfland and Hohenberg (WHH model) for a clean-limit superconductor,\cite{WH1966a,WH1966b,gurevich10a} shown as dashed lines. The parameters used in the calculations are summarized in Table\,\ref{tableHc}. For a clean-limit superconductor, orbital-limiting critical fields can be calculated with the formula $H^{orb}_{c2}\,=\,-0.73 T_c dH_{c2} / dT |_{T_c}$. Given the similarities between these values and the ones obtained with the WHH model, we conclude that the orbital pair-breaking mechanism dominates the upper critical field of \gd\ and \dy. In both compounds, the slopes of $H_{c2}$ at \tc\ increase with pressure, resulting in decreasing values of Fermi velocities $v_{\mathrm{F}}$ ($dH_{c2} / dT|_{T_c}$\,$\propto$\,$T_c / v_{\mathrm{F}}^2$),\cite{WH1966a} as shown in the inset of Fig.\,\ref{fig02}(h). As pressure increases and the volume of the unit cell decreases, enlarged cyclotron orbits at the Fermi surface would lead to enhanced Fermi velocities. However, an increase of the effective mass of the quasiparticles with pressure can compensate the effect of the expanding orbits, resulting in an effective reduction of the Fermi velocities values with increasing pressure.

Critical fields of \tb\ were also obtained at 1.75\,GPa from the ac-susceptibility measurements displayed in the inset of Fig.\,\ref{fig02}(i), as described in Section\,\ref{section2}. In this case, the voltage was measured with a SR-830 lock-in amplifier, with a sinusoidal excitation of 26\,$\mu$A and 16.6\,Hz, pre-amplifying the measured signal by a factor of 100. Calculations of \hc\ with the WHH model yield also a pure orbitally-limited superconducting state for \tb\ for $H$\,$\parallel$\,$b$. Clearly, the three critical field curves do not correlate with the monotonic decrease of lattice parameter $a$ via chemical pressure. This unexpected non-monotonic evolution of the slopes of \hc\ from \gd\ to \tb\ to \dy\ cannot be attributed to an error in the determination of the pressure or to the small difference in pressure among these measurements. We estimate that the slope of \hc\ of \dy\ at 1.5\,GPa would be similar to that of \tb\ at 1.75\,GPa.

{\renewcommand{\arraystretch}{1.2}
\begin{table}
\begin{center}
\caption{\label{tableHc} List of parameters used in the critical-field calculations shown in Fig.\,\ref{fig02}(g-i).}
\vspace{5 mm}
\begin{tabular*}{8cm}{c @{\extracolsep{\fill}} c c c c c c}
\hline
\hline
\noalign{\smallskip}
           &$P$&$T_c$& $H^{orb}_{c2}$&$\frac{dH_{c2}}{dT}|_{T_c}$&$v_F$\\        
\noalign{\smallskip}
           & GPa &K &Oe &Oe/K & m/s\\        
\noalign{\smallskip}
\hline
                  & 1.2  & 0.55 &  18 &  -45 & 197000\\
GdTe$_{3}$ & 1.8  & 0.87 &  76 & -120 & 152000\\
                  & 2.5  & 1.13 & 140 & -175 & 143000\\  
\hline
TbTe$_{3}$ & 1.75  & 0.84 &  26 & -42 & 252000\\
\hline
                   & 1.2  & 0.73 & 18 & -35 & 257000\\
DyTe$_{3}$ & 1.8  & 0.92 & 35 & -52 & 237000\\
                  & 2.5  & 1.20 & 61 & -70 & 233000\\      
\hline
\hline
\end{tabular*}
\end{center}
\end{table}

\subsection{Pressure dependence of \tc\ ($H$\,=\,0)}\label{section3.C}
The pressure dependence of the superconducting transition temperature \tc\ of \gd, \tb\ and \dy\ is plotted in Fig.\,\ref{fig03} ($H$\,=\,0). Data from Ref.\,\onlinecite{hamlin09a} have also been included. Near $P$\,=\,5\,GPa, \tc\ increases sharply from $\sim$\,1.5\,K to $\sim$\,3.5\,K. Abrupt changes in \tc\ can usually be explained in terms of structural phase transitions or multiple superconducting phases. In the first scenario, if a transformation from orthorhombic to tetragonal structure was taking place in the lattice, then the loss of broken spatial symmetry could result in the enhancement of \tc\ at higher pressures.\cite{crawford05a} Recent x-ray experiments under pressure performed on \ce\ revealed that the slight orthorhombic distortion that exists at ambient pressure is gradually suppressed as pressure is increased, until 3\,GPa where the in-plane $a$ and $c$ lattice parameters become indistinguishable.\cite{sacchetti09a} Despite the fact that the volume of the unit cell decreases smoothly with pressure, it is nevertheless possible that opposite changes of the lattice parameters compensate one another. In brief, we cannot rule out a structural phase transition as a reason for the abrupt variations in \tc\ observed at $P$\,=\,5\,GPa.

\begin{figure}[t]
\begin{center}
\includegraphics[width=3.2in]{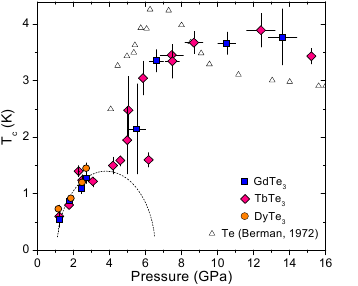}\\
\caption{(Color online) Superconducting \tc\ (onset) versus pressure $P$ phase diagram of \gd, \tb\ and \dy. Data below 2.7\,GPa were obtained in the hydrostatic clamped-cell experiments, while higher-pressure data were obtained from the Bridgman-anvil cell experiments. Data corresponding to tellurium is displayed for comparison (white triangles). The dashed line represents a dome-shaped region where the superconducting phase intrinsic to the \rte\ possibly occurs.}
\label{fig03}
\end{center}
\end{figure}

For the second scenario, given that the samples were grown via self-flux technique, one might consider that tellurium inclusions could be present in the samples. Tellurium is an insulating material at ambient pressure; Matthias and Olsen\cite{matthias64} found that it metalizes and becomes superconducting above 5\,GPa. The study of the pressure dependence of \tc\ of Te was performed by Berman, Bynzarov and Kurkin\cite{berman72} and later by Bundy and Dunn\cite{bundy80} who found that \tc\ reaches a maximum value of 4.3\,K at 6.3\,GPa in a low-pressure phase and 8\,K above 30\,GPa in a high-pressure phase.\cite{akahama92} The data from Berman \textit{et\,al.} are shown next to the \rte\ data in Fig.\,\ref{fig03} (white triangles). The superconducting phase of Te overlaps with the high-pressure SC region of the \rte, suggesting that the SC observed in the \rte\ above 5\,GPa might originate from percolative tellurium inclusions. Regarding the low-pressure SC region, recent x-ray powder-diffraction experiments revealed that a straight line with negative slope separates the insulating Te-I (hexagonal) and the metallic Te-II (monoclinic) phases (from approximately 2.8\,GPa and 730\,K to 4.4\,GPa and 290\,K),\cite{hejny06} indicating that Te should remain insulating below room temperature at pressures below 5\,GPa. Thus, additional tests on the low-$P$/low-$T$ region of the phase diagram were necessary to shed light on these matters.
\begin{figure}[t]
\begin{center}
\includegraphics[width=3.2in]{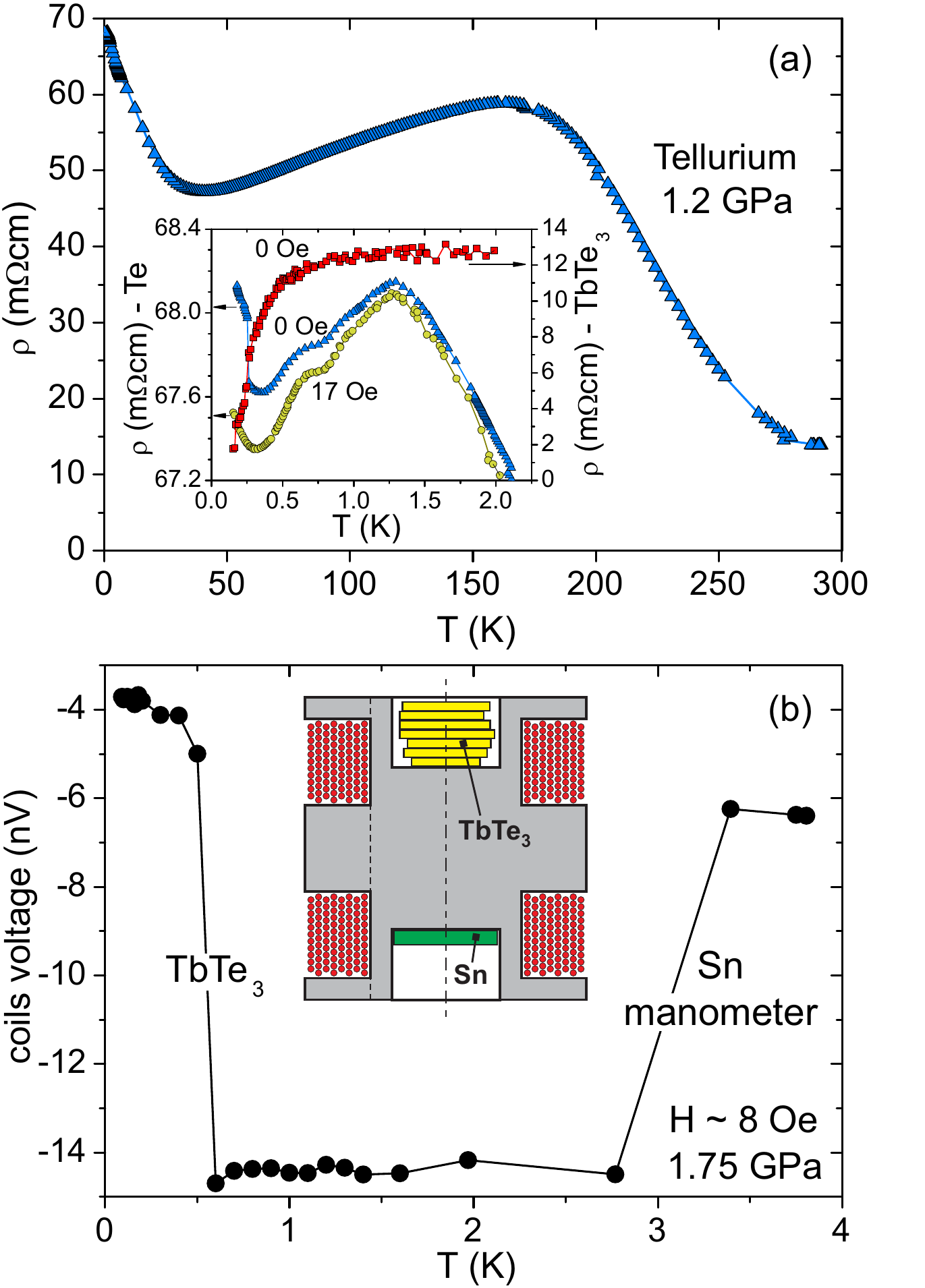}\\
\caption{(Color online) (a) Electrical resistivity $\rho$ versus temperature $T$ of tellurium at 1.2\,GPa. The sample of Te was measured in a hydrostatic-clamped cell together with a sample of \tb. At 1.2\,GPa, Te is non-metallic down to 150\,mK while \tb\ superconducts below \tc\,=\,0.7\,K (inset, $H$\,$\parallel$\,$b$). (b) Ac-susceptibility measurement of \tb\ at 1.75\,GPa ($H$\,$\parallel$\,$b$). The cartoon illustrates the secondary coils used in the high-pressure experiments.} 
\label{fig04}
\end{center}
\end{figure}

The electrical resistivity of a high-purity sample of Te was measured along with a sample of \tb\ in the hydrostatic cell at 1.2\,GPa [Fig.\,\ref{fig04}(a)]. Clearly, the Te sample is non-metallic at this pressure, and no signs of SC were found down to 150\,mK [inset of Fig.\,\ref{fig04}(a)]. In contrast, at the same pressure, \tb\ superconducts with \tc\,=\,0.7\,K, in agreement with the low-pressure region of Fig.\,\ref{fig03}. These results favor the hypothesis that the SC observed at low-$P$ and low-$T$ is not due to Te-inclusions but intrinsic to the rare-earth tritellurides. The dashed line in Fig.\,\ref{fig03} illustrates this scenario, where the superconducting phase of \rte\ forms in a dome-like region.

In order to quantify the Meissner fraction of the superconducting phase intrinsic to the \rte, we measured ac-susceptibility under pressure of \tb, as described in Section\,\ref{section2}. Several pieces of \tb\ (0.86\,mg, 0.60\,10$^{-4}$\,cm$^3$) were loaded in one side of the secondary coils and a disc of Sn (0.44\,mg, 1.13\,10$^{-4}$\,cm$^3$) on the opposite secondary coil, as it is illustrated in Fig.\,\ref{fig04}(b). The secondary coil system was then loaded in a hydrostatic cell under the same pressure conditions as described previously. The Sn disc served, on the one hand, as the pressure manometer, and on the other hand, to compare the susceptibility jump of \tb\ with the one of a well known type-I superconductor. Fig.\,\ref{fig04}(b) displays the results obtained at 1.75 GPa. Upon cooling, the voltage first drops just above 3\,K, indicative of the onset to the superconducting state of the Sn sample, and an opposite jump occurs below 1\,K, consistent with the \tc\ of \tb\ at 1.75\,GPa and 8\,Oe. Assuming that the Sn sample displayed full shielding of the magnetic field (100\%), we estimate that the SC transition of \tb\ corresponds to $\sim$ 70\% shielding. This value could be closer to 100\% if we consider that most of the TbTe$_3$ samples were located off-centered with respect to the coils.

\section{Discussion}\label{section4}
The temperature versus pressure phase diagram of \gd, \tb\ and \dy\ is presented in Fig.\,\ref{fig05}. The suppression of \tupper\ with pressure for the three compounds occurs almost linearly at a rate $\mathrm{d}T_{\mathrm{CDW,1}}/\mathrm{d}P$\,$\sim$\,$-$85\,K/GPa, comparable to the rates observed in transition-metal chalcogenides, such as NbSe$_3$ ($-$\,45\,K/GPa),\cite{manolo93} TaS$_3$ ($-$\,30\,K/GPa),\cite{manolo93} and 1$T$-TiSe$_2$ ($-$\,50\,K/GPa).\cite{kusmartseva09} Unfortunately, it was not possible to determine \tupper\ and \tlower\ from the Bridgman-anvil cell experiments ($P$\,$>$2.7\,GPa) due to the presence of pressure gradients that smear out the features in resistivity. Using the Birch-Murnaghan equation for the pressure dependence of the lattice volume,\cite{murnaghan44} one can roughly estimate the rate of suppression of \tupper\ obtained with chemical pressure. From \gd\ to \dy, \tupper\ decreases by 71\,K and the unit cell contracts by 1.5\,\% (ambient pressure).\cite{ru08a} Linear bulk modulus parameters $B_0$\,=\,59\,GPa and $B'$\,=\,5.6 were recently measured for \ce.\cite{sacchetti09a} Assuming that these values do not vary much across the \rte\ series, it follows that $\mathrm{d}T_{\mathrm{CDW,1}}/\mathrm{d}P$\,$\sim$\,$-$80\,K/GPa for chemical pressure, similar to the effects of externally applied pressure.

The lower CDW ordering temperatures \tlower\ increase with pressure from the ambient pressure values previously reported for \dy\ and \tb, while for \gd, \tlower\ can first be determined in the 1.3\,GPa resistivity curve. A quantum critical point due to the emerging lower CDW in \gd\ would likely be located below that pressure. The \tlower($P$) data appear to converge with the \tupper($P$) lines at intermediate pressures, resulting in a single phase boundary separating the high-temperature disordered state and the double-CDW state occurring at lower temperatures. The convergence of the upper and lower CDWs could be due to the loss of lattice distortion taking place upon entering a tetragonal-structure state at higher pressures. X-ray diffraction experiments under pressure performed on \ce\ showed that the $a$ and $c$ lattice parameters become identical above 3\,GPa, supporting this hypothesis.\cite{sacchetti09a} Additional studies of the lattice structure of the \rte\ compounds under pressure are still required to understand the interplay of the two CDW phases and to determine the evolution of $q_1$ and $q_2$ in the $T$-$P$ phase diagram. 
\begin{figure}[t]
\begin{center}
\includegraphics[width=3.4in]{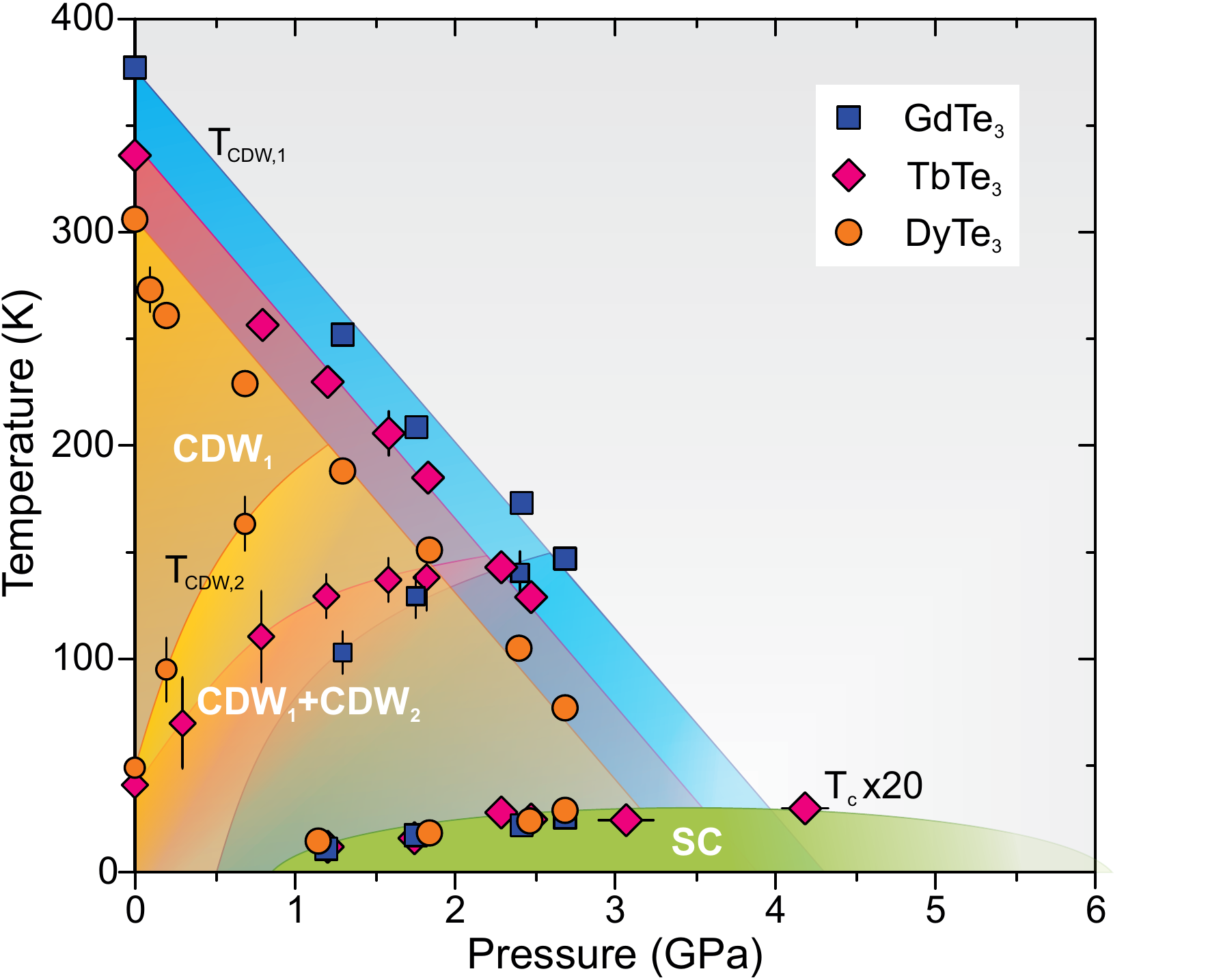}\\
\caption{(Color online) Temperature versus pressure phase diagram of \gd, \tb\ and \dy. \tupper\ is suppressed with pressure and merges with \tlower, which can no longer be determined above 2.7\,GPa. At lower temperatures, SC emerges above 1\,GPa and below 1.5\,K (\tc\ from low-pressure region of Fig. 3, amplified by a factor of 20). Ambient pressure data taken from Refs.\,\onlinecite{ru08a} and \onlinecite{banarjee13}.}
\label{fig05}
\end{center}
\end{figure}	

The results presented in Section\,\ref{section3} indicate that in the lower pressure region of the phase diagram of \gd, \tb\ and \dy, the interplay and suppression of two CDW orders tuned by pressure give rise to SC emerging above 1\,GPa and below 1.5\,K, intrinsic to the \rte\ system. Competition between CDW and SC could be understood in terms of a Bilbro-McMillan partial gaping scenario of the Fermi surface,\cite{bmcm76,hamlin09a} as previously proposed for layered transition-metal chalcogenides and $\alpha$-uranium.\cite{lander94} If competition between these phases took place, one would then expect the presence of dome-like SC regions for each \rte\ compound, centered around critical pressures \pc\ at which the CDWs were fully suppressed to zero temperature and the maximum values of \tc\ were attained.\cite{manolo15} This possibility is illustrated as the green dome-like region of Fig.\,\ref{fig05}. Despite the differences in the values of $T_{\mathrm{CDW}}$, the SC regions for the three \rte\ are indistinguishable from each other. The SC phase developing at higher pressures (not displayed in Fig.\,\ref{fig05}) impedes further analysis. X-ray diffraction data of lighter LaTe$_3$ obtained at 5 and 6 GPa revealed an upper CDW suppressed at a similar rate as for the Gd, Tb and Dy samples,\cite{sacchetti09a} suggesting that SC could appear in the La compound above 5\,GPa. For clarity, the AFM ordering temperatures \tn\ are not displayed in this phase diagram, which appear below 10\,K and show a weak pressure dependence up to 3\,GPa.

We expected to observe a de Gennes' scaling of the values of \tc\ for the different \rte\ compounds due to Abrikosov-Gor'kov (AG) pair-breaking mechanism, as it is observed in other rare-earth and transition metal alloys.\cite{maple82} This is based on an assumption, not unreasonable, that the unreconstructed electronic structures are essentially identical for the three compounds, with only subtle differences that affect their degree of nesting. In that case, once the CDWs are suppressed, the electron-phonon pairing strength would be the same in all three compounds, modified only by AG pair-breaking. However, at a same or similar pressure, we found that the values of \tc\ of \gd, \tb\ and \dy\ do not follow de Gennes' scaling, or at least, they are so close to each other that it is not possible to define a clear trend within experimental error.

It remains not clear what is the role of the rare-earth atoms on the charge-order and SC of these compounds. The pressure dependence of the magnetic ordered state of the rare-earth atoms and its interaction with the CDW has been studied so far in detail for \ce\ which shows no signs of SC.\cite{zocco09a} For \gd, \tb\ and \dy, the N\'{e}el temperatures display very little pressure dependence, so the rare-earth magnetism appears to be completely decoupled from the CDW and SC states. On the other hand, as it has been already noted in Section\,\ref{section3.C} and plotted in Fig.\,\ref{fig02}(i), the critical field curves of \gd, \tb\ and \dy\ do not correlate with the monotonic, smooth decrease of lattice parameter $a$ across the series, suggesting that the magnetism of the $R$-Te layers could play an important role in the formation of the superconducting state. For example, recent neutron scattering experiments showed that correlations leading to the long-range magnetic order in \tb\ are linked to the modulations that occur in the CDW.\cite{pfuner12} Moreover, the onset of the long-range magnetic order in \dy\ is characterized by the opening of a superzone gap observed in electrical resistivity, which could be responsible for additional reconstruction of the Fermi surface of this compound at low temperatures.\cite{ru08b}  A thorough study of the pressure dependence of the rare-earth magnetism in \rte\ ought to be performed in order to clarify these matters.

In summary, we have studied the interplay of charge-density-waves, rare-earth magnetism and superconductivity in \gd, \tb\ and \dy\ at high pressures. The suppression of the two charge-density-wave states with pressure gives rise to superconductivity at lower temperatures. The ac-susceptibility experiments on \tb\ provide compelling evidence for bulk SC in the low-pressure part of the phase diagram, intrinsic to the \rte\ compounds. Above 5\,GPa, SC with \tc\,$>$\,3\,K could be attributed to a structural phase transition or percolative inclusions of superconducting tellurium.

\subsection{ACKNOWLEDGMENTS}

D.Z. thanks M.\,N\'{u}{\~n}ez-Regueiro, F.\,Weber, R.\,Eder, R.\,Heid, R.\,Hott and H.\,v.\,L\"{o}hneysen for stimulating discussions. High-pressure research at University of California, San Diego, was supported by the National Nuclear Security Administration under the Stewardship Science Academic Alliance program through the U.S. Department of Energy Grant No.\,DE-NA0001841. Work at Stanford University was supported by the DOE, Office of Basic Energy Sciences, under Contract No.\,DE-AC02-76SF00515.


\begin{thebibliography}{48}
\bibitem[{not()}]{note1}
\bibinfo{note}{Due to difficulties arising from tuning small magnetic fields of
  the order of 1\,Oe, the highest \tc\ occurred at fields other than zero, even
  after oscillating the magnet to zero field, possibly due to an intrinsic
  offset of the magnet power supply or to remnant fields in the magnet. The
  values of magnetic field shown in Fig.\,\ref{fig02} were corrected by
  appropriate values of the order of 17\,Oe (within an error of $\pm$\,2\,Oe).}
\end{thebibliography}
\end{document}